\documentclass[aps,preprint,a4paper,showpacs,amsmath,amssymb,floatfix]{revtex4}

\usepackage{graphicx}
\usepackage{dcolumn}
\usepackage{bm}

\newcommand{\Op}[1]{{\boldsymbol{\mathrm{\hat{#1}}}}}
\newcommand{\beq}{\begin{equation}}
\newcommand{\eeq}{\end{equation}}
\newcommand{\beqar}{\begin{eqnarray}}
\newcommand{\eeqar}{\end{eqnarray}}
\newcommand{\bea}{\begin{eqnarray}}
\newcommand{\eea}{\end{eqnarray}}
\newcommand{\bcen}{\begin{center}}
\newcommand{\ecen}{\end{center}}

\begin{document}
\draft
\title{The Quantum Refrigerator: The quest for absolute zero}

\author{Yair Rezek$^{a}$, Peter Salamon$^{b}$, Karl Heinz Hoffmann$^c$, and Ronnie Kosloff$^{a}$}
\affiliation{$^{a}$Fritz Haber Research Center for Molecular Dynamics,
Hebrew University of Jerusalem, Jerusalem, 91904, Israel.\\
$^{b}$ Department of Mathematical Science, 
San Diego State University, San Diago, California, 92182, USA.\\
$^{c}$ Physics Institute,
Technical University of Chemnitz, Chemnitz, D-09107 Germany}

\email{ronnie@fh.huji.ac.il}

\pacs{05.70.Ln, 07.20.Pe}
\begin{abstract}
The scaling of the optimal cooling power of a reciprocating quantum refrigerator is sought
as a function of the cold bath temperature as $T_c \rightarrow 0$.
The working medium consists of noninteracting particles in a harmonic potential. 
Two closed-form solutions of the refrigeration cycle are analyzed, and compared to a numerical optimization scheme, focusing on cooling toward zero temperature.
The optimal cycle is characterized by linear relations between the heat extracted from the cold bath, the energy level spacing of the working medium and the temperature.
The scaling of the optimal cooling rate is found to be
proportional to $T_c^{3/2}$ giving a dynamical interpretation to the third law of thermodynamics.
\end{abstract}
\maketitle

\section{Introduction}

Walter Nernst stated the third law of thermodynamics as follows:
``it is impossible by any procedure, no matter how idealized, to reduce 
any system to the absolute zero of temperature in a finite number 
of operations'' \cite{nerst06,nerst06b}. This statement has been
termed the unattainability principle {\cite{landsberg56,landsberg89,wheeler91,belgiorno03}.
In the present study the unattainability statement is viewed 
dynamically as  the vanishing of the cooling rate $\dot {\cal Q}_c$ 
when pumping heat from a cold bath whose temperature
approaches absolute zero.  Finding   
a limiting scaling law between the rate of cooling and temperature 
$\dot {\cal Q}_c \propto T_c^{\delta}$ quantifies the unattainability principle.

The second law of thermodynamics already imposes a restriction on $\delta$ \cite{k156}.
For a cyclic process entropy is generated only in the baths:
$\sigma = -\dot {\cal Q}_c/T_c ~+~\dot {\cal Q}_h/T_h > 0$
.
If $\dot {\cal Q}_h$ stays bounded, $| \dot {\cal Q}_h |<C$, as $T_c$ approaches $0$, then rearranging the inequality above gives 
$ C/T_h >   \dot {\cal Q}_h/T_h >  \dot  {\cal Q}_c/T_c $,
and so: $\left( \frac{ C}{T_h}\right)T_c  >  \dot  {\cal Q}_c $.
This forces $ \dot {\cal Q}_c \rightarrow 0$ as $T_c \rightarrow 0$ and, expanding $  {\cal Q}_c$ as a series near $T_c=0$, the dominant power $\delta$ in $ \dot {\cal Q}_c\propto T^{\delta}$ 
must satisfy $\delta ~\ge~ 1$.  Such an
exponent has been realized in  refrigerator models \cite{k152,k156} where the source
of irreversibility is the heat transfer. The vanishing of $\dot {\cal Q}_c$ 
is also consistent with the vanishing of the quantum unit of heat transport 
$\frac{\pi^2 k_B^2 T_c}{3 \hbar}$ \cite{rego}. Our goal in the present study is to set more stringent limits on the exponent $\delta$ for a reciprocating four stroke cooling cycle.  The cooling rate is replaced by the average refrigeration power 
${\cal R}_c={\cal Q}_c/\tau$ where $\tau$ is the cycle period.
\begin{figure}
\vspace{0.3cm}
\center{\includegraphics[width=0.6\textwidth]{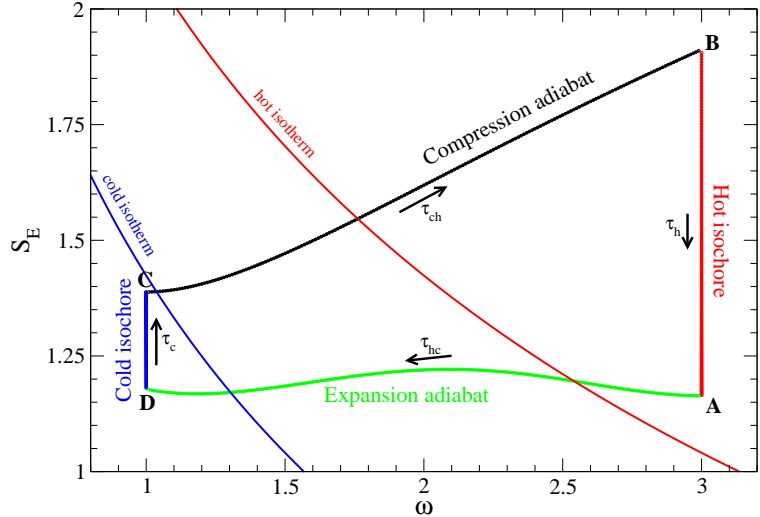}}
\caption{\label{fig:1} A typical optimal cooling power cycle ADCB
in the energy entropy ${\cal S}_E$ and frequency $\omega$ plane.}
\label{fig1}
\end{figure}

\section{The quantum Otto heat pump}

We consider a refrigerator using a controllable quantum medium as its working fluid. Our objective is to optimize the cooling rate in the limit when the temperature $T_c$ of the cold bath approaches absolute zero. A necessary condition for operation is that upon contact with the cold bath
the temperature of the working medium be lower than the bath temperature $T_{int} \le T_c$ \cite{mahler08}. The opposite condition exists on the hot bath. To fulfill these requirements the external controls modify the internal temperature by changing the energy level spacings of the working fluid. The control field varies between two extreme values $\omega_c$ and $\omega_h$, 
where $\omega$ is a working medium frequency induced by the external field.  
The working medium consists of 
an ensemble of non interacting particles in a harmonic potential. The Hamiltonian of this system,
$\Op H=\frac{1}{2m}{\Op P^2}+\frac{K(t)}{2}{\Op Q^2}$, is controlled by changing the curvature $K=m\omega^2$ of
the confining potential. 

The cooling cycle consists of two heat exchange branches alternating with two adiabatic branches (Cf. Fig. 1). The heat exchange branches (the {\em isochores}) take place with $\omega=$constant, while the adiabatic branches take place with the working medium decoupled from the baths. This is reminiscent of the Otto cycle in which heat  is transfered to the working medium  from the hot and cold baths under constant volume conditions. 

The heat carrying capacity of the working medium limits the amount of heat ${\cal Q}_c$
which can be extracted from the cold bath:
\begin{equation}
{\cal Q}_c ~~=~~ E_C -E_D ~~=~~ \hbar \omega_c( n_C- n_D )
\label{eq:fr}
\end{equation}
where $E_{C} $ is the working medium internal energy at point $\bf C$ (Fig. \ref{fig1}),
$E_D$ is the energy at point $\bf D$ and $n=\langle \Op N \rangle $ is the expectation value of the number operator. 
Examining Fig.  \ref{fig1} $n_{C} \le n_C^{eq}$ and $n_D \ge  n_A$, where equality
is obtained under the quantum adiabatic condition \cite{ft12,Kato1950}. This means also $ n_D \ge n_h^{eq}$,  leading to
${\cal Q}_c ~~\le~~ \hbar \omega_c \left( n_c^{eq} ~-~ n_h^{eq} \right)$.
Maximum ${\cal Q}_c$
is obtained for high frequency $\hbar \omega_h \gg {k_B T_h} $, leading to 
$n_h^{eq} = 0$ and $E_A=\frac{1}{2}\hbar \omega_h$ being the ground state energy. Then
for $T_c \rightarrow 0$:
\begin{equation}
{\cal Q}_c^* ~~=~~\hbar \omega_c  n_c^{eq}~~=~~\hbar \omega_c e^{- \frac{\hbar \omega_c}{k_B T_c}}      ~\le~ k_B T_c
\label{eq:qc}
\end{equation}
where we have substituted the value of $n_c^{eq}$ obtained from the partition function and the last inequality is obtained by optimizing with respect to $\omega_c$
leading to $\hbar \omega_c^* = k_B T_c$. The general result is that as $T_c \rightarrow 0$, ${\cal Q}_c^*$ and $\omega_c^*$ become 
linear in $T_c$. 

Only a  finite cycle period $\tau$ leads to a non vanishing cooling power ${\cal R}_c={\cal Q}_c/\tau$ \cite{PhysToday}. This cycle time $\tau=\tau_{hc}+\tau_c+\tau_{ch}+\tau_h$ is the sum of the times allocated to each branch Cf. Fig. \ref{fig1}. 
An upper bound on the cooling rate $ {\cal R}_c$ is required to limit the exponent as $T_c \rightarrow 0$. The optimal cooling rate $ {\cal R}_c^{opt}$ depends on the time allocated to the different branches.

The dynamics on the {\em adiabatic} segments is generated by an externally 
driven time dependent Hamiltonian $\Op H (\omega(t))$. 
The equation of motion for an operator $\Op O$ of the working medium is:
\begin{equation}
 \frac {d \Op O(t) }{dt}= \frac{i}{\hbar}[\Op H(t) ,\Op O(t)] +\frac {\partial \Op O(t) }{\partial t}
 ~~.
\label{eq:heisenberg}
\end{equation}
Typically $[\Op H(t),\Op H(t')] \neq 0$ which leads to friction like phenomena \cite{k176,k221}.
Too fast adiabatic segments will generate parasitic internal energy which will have to be dissipated
to the heat baths, thus limiting the performance.
The external power of the compression/expansion
segments is the rate  of change of the internal energy of the working medium \cite{k24}.
Therefore inserting $\Op H$ for $\Op O$ in Eq. (\ref{eq:heisenberg}) leads to the power 
$ \frac{d E}{dt} ={\cal P}~=~\langle\frac{ \partial \Op H}{\partial t} \rangle $.
The dynamics on the {\em adiabatic } segments is unitary, therefore the von Neumann entropy 
${\cal S}_{vn}=-k_B tr \{ \Op \rho \ln \Op \rho \}$ is constant. In contrast the energy entropy ${\cal S}_{E}$
changes, where ${\cal S}_{E}=-k_B \sum_j P_j \ln P_j $  and $P_j= tr \{ |j \rangle \langle j | \Op \rho \}$ 
is the probability of occupying the energy level $j$. Constant ${\cal S}_E$ is obtained 
only under quasistatic conditions.

Using the Heisenberg picture, the dynamics on the heat exchange branches, termed {\em isochores}, are generated by
${\cal L^*}(\Op O) = \frac{i}{\hbar}[\Op H,\Op O] +{\cal L}_D^*(\Op O)$ \cite{lindblad76}
with the dissipative Lindblad term ${\cal L}_D^*$ leading the system toward thermal equilibrium of an harmonic oscillator defined by
$\frac{k_{\uparrow}}{k_{\downarrow}}=\exp(-\frac{\hbar \omega}{k_B T})$ \cite{k221}. 
For the dissipative dynamics, the heat flow from the cold/hot bath is 
$ \dot {\cal Q} ~~=~~\langle{\cal L}_D ( \Op H ) \rangle$ \cite{k176,k221}.

At thermal equilibrium the energy expectation value is sufficient to fully characterize the state of 
a system. For the  working medium not in equilibrium, there is a family of generalized Gibbs states \cite{k221} that completely characterize the system during the cycle. These Gibbs states are
defined by three operators: the time dependent Hamiltonian 
$\Op H=\frac{1}{2m}{\Op P^2}+\frac{K(t)}{2}{\Op Q^2}$, 
the Lagrangian $\Op L = \frac{1}{2m}{\Op P^2}-\frac{K(t)}{2}{\Op Q^2}$ 
and the correlation  $\Op C = \omega(t)\frac{1}{2}( \Op Q \Op P +\Op P \Op Q ) $. As a result $\Op \rho =\Op \rho(\Op H, \Op L, \Op C)$.
Starting from any initial state, the system will reach a  generalized Gibbs state and will remain in such states. 
The invariance of the set ${\Op H, \Op L, \Op C}$ under the equation of motion, is due to this set forming a closed Lie algebra, 
which leads to closed equations of motion on the 
{\em adiabats} as well as on the {\em isochores} \cite{k221,k190}. 

The dynamics of the operators on the {\em adiabats} 
is obtained from Eq. (\ref{eq:heisenberg}):
\begin{equation}
\frac{d}{dt}\left(\begin{array}{c}
\Op H\\
\Op L\\
\Op C
\end{array}\right)(t)~=~ \omega(t)
\left(\begin{array}{ccc}
\mu & -\mu & 0\\
-\mu & \mu & -2\\
0 & 2  & \mu
\end{array}\right)
\left(\begin{array}{c}
\Op H\\
\Op L\\
\Op C
\end{array}\right)(t)~,
\label{eq:adiabatdy1}
\end{equation}
where $\mu =\frac{\dot \omega}{\omega^2}$ is the dimensionless adiabatic parameter.
The power becomes:
${\cal P} = \mu \omega (\langle \Op H \rangle - \langle \Op L \rangle)$ \cite{k190,k221}.
The solution of Eq. (\ref{eq:adiabatdy1}) depends on the functional form of $\omega(t)$.
When $\mu \ll 1$,
the number $n(t)$ will remain constant on the {\em adiabats}; these are the quasistatic conditions. 
For most other functions $\omega(t)$, the time evolution will involve some quantum friction \cite{k221} 
and $n_f \ge n_i$ due to the resultant parasitic 
increase in the internal energy $\Delta E=\hbar \omega_f (n_f-n_i)$.
The dissipation of this energy in particular into the cold bath counters the cooling:
${\cal Q}_c \le \hbar \omega_c (n_c^{eq}-n_c) $, therefore when $n_c > n_c^{eq}$ the refrigerator 
can no longer cool.

On the {\em isochores} the energy displays an exponential approach to equilibrium:
\begin{equation}
\frac{d \Op H}{dt} ~~=~~ -\Gamma ( \Op H - \langle \Op H \rangle_{eq} \Op I )
\label{eq:hamrelax}
\end{equation}
where $\Gamma=k_{\downarrow}-k_{\uparrow}$ is the heat conductance.
$\langle \Op H \rangle_{eq}$ is the equilibrium
expectation of the energy. The heat transfer becomes:  
$ \dot {\cal Q}=-\Gamma ( \langle \Op H \rangle -\langle \Op H \rangle_{eq}) $.

The operators $\Op L$ and $\Op C$ display an oscillatory decay to an expectation
value of zero at equilibrium:
\begin{eqnarray}
\frac{d}{dt}\left(\begin{array}{c}
\Op L\\
\Op C\\
\end{array}\right)(t)=
\left(\begin{array}{cc}
 -\Gamma & -2\omega  \\
 2\omega  & -\Gamma \\
\end{array}\right)
\left(\begin{array}{c}
\Op L\\
\Op C\\
\end{array}\right)(t)
\label{eq:motion}
\end{eqnarray}
The equation of motion (\ref{eq:adiabatdy1}), (\ref{eq:hamrelax}) and (\ref{eq:motion})
can be solved in closed form for certain special choices of $\omega(t)$ (Cf. section \ref{OptimSection} below) and numerically for any given functions $\omega(t)$ and time allocation to the branches.
After a few cycles, the refrigerator settles down to a periodic limit cycle \cite{k201}, which allows to calculate
the cooling power ${\cal R}_c={\cal Q}_c /\tau$ from the expectations of
$\Op H, \Op L, \Op C$ in the limit cycle. 

\section{Optimization of the cooling rate}
\label{OptimSection}

For sufficiently low $T_c$, the rate limiting branch of our cycle is cooling the working medium to a temperature below $T_c$ ({\bf A}$\rightarrow${\bf D} along the expansion {\em adiabat}).  As $T_c \rightarrow 0$, the total cycle time
$\tau$ is of the order of the time of this cooling adiabat, $ \tau_{hc}$, which tends to infinity.

Quantum friction is completely eliminated if the {\em adiabat} proceeds quasistatically with $\mu \ll 1$. This leads to a scaling law  ${\cal R}_c \propto T^{\delta}$ with $\delta \ge 3$. It turns out however that it is not the only frictionless way to reach the final state at energy $E_D=(\omega_c/\omega_h) E_A$. We describe two other possibilities which require less time and result in improved scaling, $\delta=2$ and $\delta=3/2$ respectively.

The first frictionless solution to Eq. (\ref{eq:adiabatdy1}) is obtained for $\mu=const$, by changing the time variable 
to $\theta=\int_0^t\omega(t') dt'$. Then factoring out the term $\mu \vec {\bf 1}$ and diagonalizing the 
time independent part with the eigenvalues $\lambda_0=0$ and $\lambda_{\pm}=\pm \Omega$
where $\Omega=\sqrt{\mu^2-4}$ leads to the adiabatic propagator ${\cal U}_a$ of $\Op H, \Op L, \Op C$:
\begin{equation}
{\cal U}_a(t)= \frac{\omega(t)}{\omega(0)\Omega^2}
\left(\begin{array}{ccc}
\mu^2 c-4 & \mu \Omega s& 2 \mu(c-1)\\
\mu \Omega s & \Omega^2 c & 2 \Omega s\\
-2 \mu(c-1)& -2 \Omega s  & \mu^2 -4 c
\end{array}\right)~.
\label{eq:adiabatdy}
\end{equation}
where $c =\cosh(\Omega \theta)$, $s =\sinh(\Omega \theta)$ and 
$\theta(t)=-\log(\frac{\omega(0)}{\omega(t)})/\mu$. The cycle propagator 
becomes the product of the segment propagators ${\cal U}_{cyc}= {\cal U}_{c}{\cal U}_{ch}{\cal U}_{h}{\cal U}_{hc}$, where ${\cal U}_{h/c} $ is obtained from  Eq. (\ref{eq:hamrelax}) and Eq. (\ref{eq:motion})
on the {\em isochores}.

The energy change on the expansion {\em adiabat} is the key for the optimal solution:
$A \rightarrow D$ :
\begin{equation}
E_D = \frac{1}{2}\hbar \omega_c ~\frac{1}{\Omega^2}\left(\mu^2 \cosh ( \Omega \theta_c) -4\right)
~,~\theta_c=-\frac{1}{\mu}\log\left({\cal C}\right)
\label{eq:hadi}
\end{equation}
where ${\cal C}=\frac{\omega_h}{\omega_c}$ is the compression ratio
and equilibration is assumed at the end of the hot {\em isochore} $E_A=\frac{1}{2}\hbar \omega_h$
for $\omega_h \rightarrow \infty$.
For very fast expansion $\mu \rightarrow \infty $,
$E_D = \frac{1}{4}\hbar \omega_c (1/{\cal C}+{\cal C})$. As $T_c \rightarrow 0$, $E_D=\frac{1}{4} \hbar \omega_h$ which becomes larger than $E_c^{eq}$ therefore the cooling stops due to friction. For the limit of infinite time $\mu \rightarrow 0$ leading to the frictionless result characterized by 
constant $n$ and ${\cal S}_E$. Then
$E_D \rightarrow \frac{1}{2}\hbar \omega_c$ which is the ground state of the oscillator. 
At this limit since $\tau \rightarrow \infty$, ${\cal R}_c=0$.
The surprising point is that we can find an additional frictionless point where $n_c=n_h$,
when $\cosh (\Omega \theta_c)=1$. Then  $\mu < 2$ and $\Omega$ becomes imaginary leading to the critical points:
\begin{eqnarray}
\mu^* = -\frac{2 \log \left({\cal C}\right)}{\sqrt{4 \pi ^2 + \log \left({\cal C}\right)^2 }} \label{eq:mucrit}
\end{eqnarray}
and $\tau_{hc}^* = (1-{\cal C})/ (\mu^* \omega_h)$.
Asymptotically as $T_c \rightarrow 0$ and $\omega_c \rightarrow 0$, the critical terms 
approach $\mu^* \rightarrow -2$ and with it the time allocation $\tau_{hc}^* ~=~ \frac{ 1}{2} \omega_c^{-1}$.
This frictionless solution with a minimum time allocation $\tau_{hc}^*$ scales as the inverse frequency $\omega_c^{-1}$ which is better than the quasistatic limit where $\tau_{hc} \propto \omega_c^{-2}$.  As we will see, it leads to $\delta = 2$.

Inspired by these findings, we sought the minimum time frictionless solution. The resulting optimal control problem \cite{peter08} is solvable leading to a second closed form solution. The optimal trajectory is of the bang-bang form with three jumps
\begin{eqnarray}
\omega(t) = \left\{
\begin{array}{l}
\omega_h,~for~t=0 \\
\omega_c, ~for~ 0 < t \le \tau_1 \\
\omega_h, ~for~ \tau_1 < t < \tau_{hc} \\
\omega_c, ~for~ t=\tau_{hc}~~,
\end{array}
\right.
\end{eqnarray}
where $\tau_1+\tau_2=\tau_{hc}$ and the times 
$\tau_1 = \frac{1}{2\omega_c} \arccos \left( \frac{\omega_h^2+\omega_c^2}{(\omega_h+\omega_c)^2} \right)$
and 
$\tau_2 = \frac{1}{2 \omega_h} \arccos \left( \frac{\omega_h^2+\omega_c^2}{(\omega_h+\omega_c)^2} \right)$
are chosen such that the number operator is preserved $n_f=n_i$.
The minimum time allocation for $\omega_c \rightarrow 0$ which is appropriate for $T_c \rightarrow 0$
becomes $\tau_{hc}^*=\frac{1}{\sqrt{\omega_h}} \omega_c^{-\frac{1}{2}}$,
which is better than the solution in Eq. (\ref{eq:taucrit}). As we show below, it leads to $\delta = 3/2$.

Both frictionless solutions lead to an upper bound on the optimal cooling rate of the form:
\begin{equation}
{\cal R}_c \le A \omega^{\nu} n_c^{eq}
\end{equation}
where $A$ is a constant and the exponent $\nu$ is either $\nu=2$ for the $\mu=const$ solution or $ \nu=\frac{3}{2}$
for the three-jump solution.
Optimizing ${\cal R}_c$ with respect to $\omega_c$ leads to a linear relation between $\omega_c$ and $T_c$, 
$\hbar \omega_c = \kappa k_B T_c$. The constant $\kappa =2+{\cal P}(-2e^{-2})\approx 1.6$ for $\nu=2$ and 
$\kappa =3/2+{\cal P}(-3/2 e^{-3/2})\approx 0.87$ for $\nu=\frac{3}{2}$, where ${\cal P}$ is the product-log function.

Once the time allocation on the {\em adiabats} is set the
time allocation on the {\em isochores} is optimized using the method of Ref. \cite{k221}:
\begin{equation}
{\cal R}_c^* ~~=~~ \frac{e^z}{(1+e^z)^2}\Gamma \hbar \omega_c (n_c^{eq}-n_h^{eq})
\end{equation}
where $z=\Gamma_h \tau_h=\Gamma_c \tau_c$ and $z$ is determined by the equation 
$2z+\Gamma (\tau_{hc}+\tau_{ch})=2 \sinh(z)$. For the limit $T_c \rightarrow 0$, $\Gamma \tau_{hc}$ is large
therefore $z$ is large leading to:
\begin{equation}
{\cal R}_c^* ~\approx~ \frac{\Gamma (\tau_{hc}+\tau_{ch})}{(1+\Gamma (\tau_{hc}+\tau_{ch}))^2} \Gamma \hbar \omega_c (n_c^{eq}-n_h^{eq})
\end{equation}
At high compression ratio $\omega_h \gg \omega_c$ and if in addition $\omega_c \ll \Gamma$ we obtain:
\begin{equation}
{\cal R}_c^* ~\approx~  \hbar \omega_c^2 n_c^{eq}
\label{eq:rate1}
\end{equation}
for the $\mu=const$ frictionless solution, and 
\begin{equation}
{\cal R}_c^* ~\approx~  \frac{1}{2}\hbar \omega_c^{\frac{3}{2}} \sqrt{\omega_h} n_c^{eq}~~,
\label{eq:rate2}
\end{equation}
for the three-jump frictionless solution.
Due to the linear relation between $\omega_c$ and $T_c$, Eq. (\ref{eq:rate1}) and (\ref{eq:rate2}) the exponent $\delta$ where $\delta=3$ for the quasistatic scheduling,
$\delta=2 $ for the constant $\mu$ frictionless scheduling and $\delta=\frac{3}{2}$ for the three-jump frictionless scheduling.

To check the optimization assumptions a numerical procedure was applied to maximize the cooling rate by adjusting the \emph{times on the four branches} for a given a choice of scheduling function and the external constraints on the cycle. These constraints are the coupling $\Gamma$, the temperatures $T_c$ and $T_h$, and the frequencies $\omega_c$ and $\omega_h$. The cooling rate optimizations employed random time allocations to the different cycle segments augmented by a guided-search algorithm. 
The choice of scheduling function $\omega(t)$  determines the exponent of the scaling in ${\cal R}_c \propto T^{\delta}$. The optimal cooling rate for  linear and exponential scheduling functions are shown in Fig. \ref{fig3}. As a final numerical corroboration, we tried a multistep genetic algorithm allowing piecewise variation of 
$\omega (t)$. The algorithm converged to a cooling rate very close to the optimal three jump solution.

Two main observations have led to the optimal exponents as $T_c \rightarrow 0$, 
the first is that the time allocation on the expansion {\em adiabat} sets the scaling 
and the second is that the frictionless cycles have superior performance. Fig. \ref{fig3} also shows the results of numerical optimizations for the two frictionless schedules. At low temperatures the time allocated to the {\em adiabats} dominates and scales as $\tau_{hc}^* \propto 1/T_c$ for the $\mu=const.$  schedule and $\tau_{hc}^* \propto 1/T_c^{1/2}$ for the three-jump schedule. Since ${\cal Q}_c$ for all cases is linear with $T_c$, the asymptotic cooling rate approaches 
$ {\cal R}_c \propto T_c^{2}$ and $ {\cal R}_c \propto T_c^{3/2}$ respectively. 

\begin{figure}
\vspace{0.3cm}
\center{\includegraphics[height=7cm]{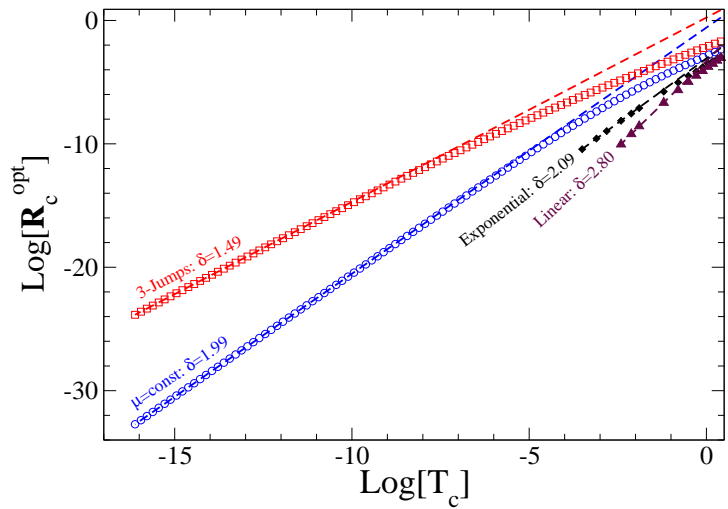}}
\vspace{0.1cm}

\caption{The cooling rate as a function of temperature for different scheduling functions $\omega(t)$. Straight lines are linear continuations of the last points. The exponent of ${\cal R}_c \propto T_c^{\delta}$ is indicated. The lowest exponent (red squares) shows 
the three-jump frictionless optimization. The next lowest (blue circles) show the result of the optimization with $\mu=const$. 
The black diamonds correspond to $\omega(t) \propto \exp( \alpha t) $  and the highest exponent
corresponds to $\omega(t) \propto t$ (magenta triangles). }
\label{fig3}
\end{figure}

\section{Discussion and Conclusion}

The optimal quantum refrigerator in the quest to reach the absolute zero temperature
shows a linear scaling of ${\cal Q}_c^*$ with $\omega_c$ and $T_c$. 
This scaling is the minimum to eliminate the divergence of the entropy generated on the cold bath.
If the energy level spacing $\hbar \omega_c$ cannot follow $T_c$ the refrigerator will be limited by a minimum temperature \cite{tova08}. If the level spacing follows $T_c$, the scaling of the cycle time is dominated by
the scheduling function $\omega(t)$ on the {\em adiabats}. 
The best results were obtained for the three-jump frictionless solutions 
which give $\tau ~\propto {\omega_c}^{-1/2} $. 
The three-jump scheduling is the minimum time frictionless solution \cite{peter08}. We conjecture that \emph{any} cooling cycle is limited by the {\em adiabatic } expansion
\cite{wheeler91}. Our conjecture implies that the unattainability principle is a consequence of dynamical considerations and is limited by the exponent ${\cal R}_c^{opt} \propto T_c^{3/2}$.

\section*{Aknowledgements}
We want to thank Tova Feldmann for support and crucial discussions.
This work was supported
by the Israel Science Foundation, The Fritz Haber center is supported by the Minerva
Gesellschaft f\"ur die Forschung, GmbH M\"unchen, Germany. PS gratefully acknowledges the hospitalities of the Hebrew University of Jerusalem and the Technical University of Chemnitz.


\end{document}